\documentclass[review]{elsarticle}

\usepackage{lineno,hyperref}
\usepackage{underscore}
\usepackage{listings}
\modulolinenumbers[5]

\journal{Journal of Parallel and Distributed Computing}

\PassOptionsToPackage{unicode}{hyperref}
\PassOptionsToPackage{naturalnames}{hyperref}
\begin{document}

\begin{frontmatter}

\title{ExpoCloud: a Framework for Time and Budget-Effective Parameter Space Explorations Using a Cloud Compute Engine}

\author{Meir Goldenberg\fnref{myfootnote}}
\address{Jerusalem College of Technology}
\fntext[myfootnote]{Email address: mgoldenb@g.jct.ac.il}

\begin{abstract}
Large parameter space explorations are among the most time consuming yet critically important tasks in many fields of modern research. ExpoCloud enables the researcher to harness cloud compute resources to achieve time and budget-effective large-scale concurrent parameter space explorations. 

ExpoCloud enables maximal possible levels of concurrency by creating compute instances on-the-fly, saves money by terminating unneeded instances, provides a mechanism for saving both time and money by avoiding the exploration of parameter settings that are as hard or harder than the parameter settings whose exploration timed out. Effective fault tolerance mechanisms make ExpoCloud suitable for large experiments.

ExpoCloud provides an interface that allows its use under various cloud environments. As a proof of concept, we implemented a class supporting the Google Compute Engine (GCE). We also implemented a class that simulates a cloud environment on the local machine, thereby facilitating further development of ExpoCloud.

The article describes ExpoCloud's features and provides a usage example. The software is well documented and is available under the MIT license \cite{ExpoCloud, ExpoCloud-devdocs}.  
\end{abstract}

\begin{keyword}
parameter space exploration\sep distributed computing\sep cloud compute engine\sep large-scale.
\end{keyword}

\end{frontmatter}

%\linenumbers

\section*{Introduction}
%% State the objectives of the work and provide an adequate background, avoiding a detailed literature survey or a summary of the results.
Large parameter space explorations are among the most time consuming yet critically important tasks in many fields of modern research. For example, computer science research is often concerned with the study of algorithms for solving computational problems, whereby the algorithm's behavior and the computation time for solving the problem are controlled by a number of parameters. The possible settings of these parameters form a large parameter space, whose thorough exploration requires that the algorithm be run to solve a number of problem instances for each parameter setting of both the algorithm and the problem. It is our assumption in this work that individual parameter settings can be explored concurrently and independently of each other.

In the absence of a tool that makes large-scale parameter explorations easy to accomplish, researchers resort to running ad hoc scripts either on a local machine or on a cluster. Most recently, cloud-based compute engines became a budget-friendly option. The amount of computational power available through such services is usually much greater than that available in the research clusters. However, the amount of technical expertise and scripting required to set up an experiment that harnesses these resources may prove to be a stumbling block. As a result, the researchers adopt simplifying limitations, such as using multiple threads on a single compute instance~\cite{Pollack}.

\subsection*{The vision}
We envisioned a framework that would let the researcher define his/her {\em workload} and achieve {\em maximal concurrency} while economizing on his/her {\em time} and {\em money}, allowing flexibility in choosing the {\em cloud platform} and providing {\em fault tolerance}. 

ExpoCloud \cite{ExpoCloud} is our implementation of the above vision. It realizes the words that appear above in italics as follows:
\begin{itemize}
    \item {\em The workload} is a list of tasks, each defined by a setting of parameters. It is computed at the commencement of the experiment and passed to the framework for automated execution.
    \item {\em Maximal concurrency} is achieved by creating a new compute instance as often as is allowed by the cloud platform, for as long as there are tasks to assign.
    \item {\em Economizing on time} is achieved by letting the user specify a deadline and a {\em hardness} (defined below) for each task. When a task takes more time to execute than the time specified by the deadline, we say that the task has {\em timed out}. ExpoCloud terminates timed out tasks automatically. 
    
    A task's {\em hardness} is a tuple of parameter values that correlate with the time required to execute the task. The researcher specifies which subset of parameters determines a task's hardness and provides a method that compares hardnesses of two tasks. When a task times out, the framework terminates all currently running tasks that are as hard or harder than the timed out task. It also avoids running such tasks in the future. The framework executes the tasks in the order from the easiest to the hardest, so as to maximize the number of tasks that do not have to be executed.
    \item {\em Economizing on money} is achieved by deleting a compute instance as soon as it is done with the tasks assigned to it and there are no more tasks to be assigned.
    \item ExpoCloud provides great flexibility for choosing the {\em cloud platform}. To adapt to a given cloud platform, one needs to merely provide an extension class with methods to create, terminate and list compute instances. In addition, the researcher is in full control of the properties of the compute instances, since all of them are created based on machine images specified by the researcher.
    \item {Fault tolerance} means that the computation would proceed even if one or more compute instances fail for any reason.
\end{itemize}

\vspace{0.1in}
Before moving on to the main part of the article, we introduce an example that we will use to demonstrate the framework's design and usage. 

\subsection*{The example parameter exploration}
Consider the agent assignment problem, as follows. A team of $n$ agents needs to complete a project consisting of $m$ tasks, where $n\ge m$. The tasks have to be performed sequentially. For each agent $i$ and task $j$, we are given $t_{ij}$, the amount of time, in seconds, that the agent $i$ requires to complete the task $j$. The problem is to assign an agent to each task, such that no agent is assigned to more than one task and the total time of completing the project is minimized.

Suppose we use the classical branch and bound (B\&B) search algorithm for solving this problem, as follows. The algorithm is recursive. It starts with an empty assignment, whereby no agent is assigned to a task. At each recursive call, it extends the current {\em partial assignment}  by assigning an agent to the next task. When all tasks have been assigned an agent, we say that the assignment is {\em full}. At this base case of the recursion, the algorithm updates the currently best full assignment and the corresponding time for completing the project. 

The advantage of B\&B search over the brute-force search is the so called B\&B {\em cutoff}. Namely, whenever the time corresponding to the current partial assignment is greater than that of the best full assignment, the current assignment can be discarded without losing the solution optimality.

A more efficient version of this algorithm uses a heuristic. Given a partial assignment, the heuristic is a lower bound on the time needed to complete the remaining tasks. This bound is computed by assigning the best of the unused agents to each of the remaining tasks, while allowing the assignment of the same remaining agent to more than one remaining task. Whenever the sum of the time corresponding to the current partial assignment and the heuristic is greater than that of the best full assignment, the current assignment can be discarded.

Thus, we have three algorithmic variants - the brute-force search, the classical B\&B search and the B\&B search with a heuristic. To thoroughly understand the properties of the agent assignment problem and the B\&B search's performance for solving it, we need to run each algorithmic variant to solve a number of generated problem instances for many possible settings of the number of agents $n$ and the number of tasks $m$.

What range of values should we consider for the number of agents $n $ and the number of tasks $m$? Without a framework like ExpoCloud, this question is not easily answered. First, the range will depend on the algorithmic variant. The brute-force search will only be able to handle small problems, while B\&B with a heuristic might be able so solve much larger instances. Knowing his/her budget of time for the whole experiment, the researcher might decide on a deadline per problem instance. He/she might then perform several test runs to get a feeling for how much time each algorithmic variant takes to solve problem instances of various sizes. Even after this laborious tuning stage, the researcher will still run the risk of some instances taking disproportionately long time, possibly stumbling the whole experiment.

With ExpoCloud, the question is really easy. First the researcher writes a short class that defines a {\em task}  as running one algorithmic variant to solve a single problem instance for one particular setting of $n$ and $m$. After deciding on a deadline, the researcher picks a large range of values for $n$ and $m$, with the upper bounds that for sure cannot be solved by the best algorithmic variant. He/she writes a single nested loop to generate all the tasks. All of this is shown in the next section.

Next, the researcher notices that larger values of $n$ correspond to harder problem instances and so do larger values of $m$. It is also clear that the same instance is likely to be solved faster by the B\&B search with a heuristic than with the classical B\&B, which is in turn faster than the brute-force algorithm. The researcher defines several short methods informing the framework of these observations and off the experiment goes. ExpoCloud will care both for stopping a problem instance as soon as it times out and for not attempting exploring parameter settings that are as hard or harder.

The researcher does not need to worry about deciding on the number of compute instances and creating those instances. Neither does he need to worry about stopping compute instances when the experiment is done. The results are easily obtained in a nice tabular format, which is again specified with a few short methods.

If the researcher wants to run the experiment locally, e.g. on his/her laptop, he/she can do that as well. ExpoCloud makes it easy to use as many CPUs of the researcher's machine as desired. Furthermore, this run is actually a simulation of performing the experiment on the cloud. It is thus a powerful tool to facilitate further development of the framework.

\vspace{0.1in}
ExpoCloud is written in Python and is available on GitHub under the MIT license~\cite{ExpoCloud}. The GitHub page contains the user documentation and links to the developer's documentation, where the source code is described~\cite{ExpoCloud-devdocs}.

The next section details the features of the framework and shows in full how the above example experiment is setup and run.

\section*{Material and methods}
%% Provide sufficient details to allow the work to be reproduced by an independent researcher. Methods that are already published should be summarized, and indicated by a reference. If quoting directly from a previously published method, use quotation marks and also cite the source. Any modifications to existing methods should also be described.

\subsection*{The overall architecture}
\begin{figure}[htp]
    \centering
    \includegraphics[width=0.9\textwidth]{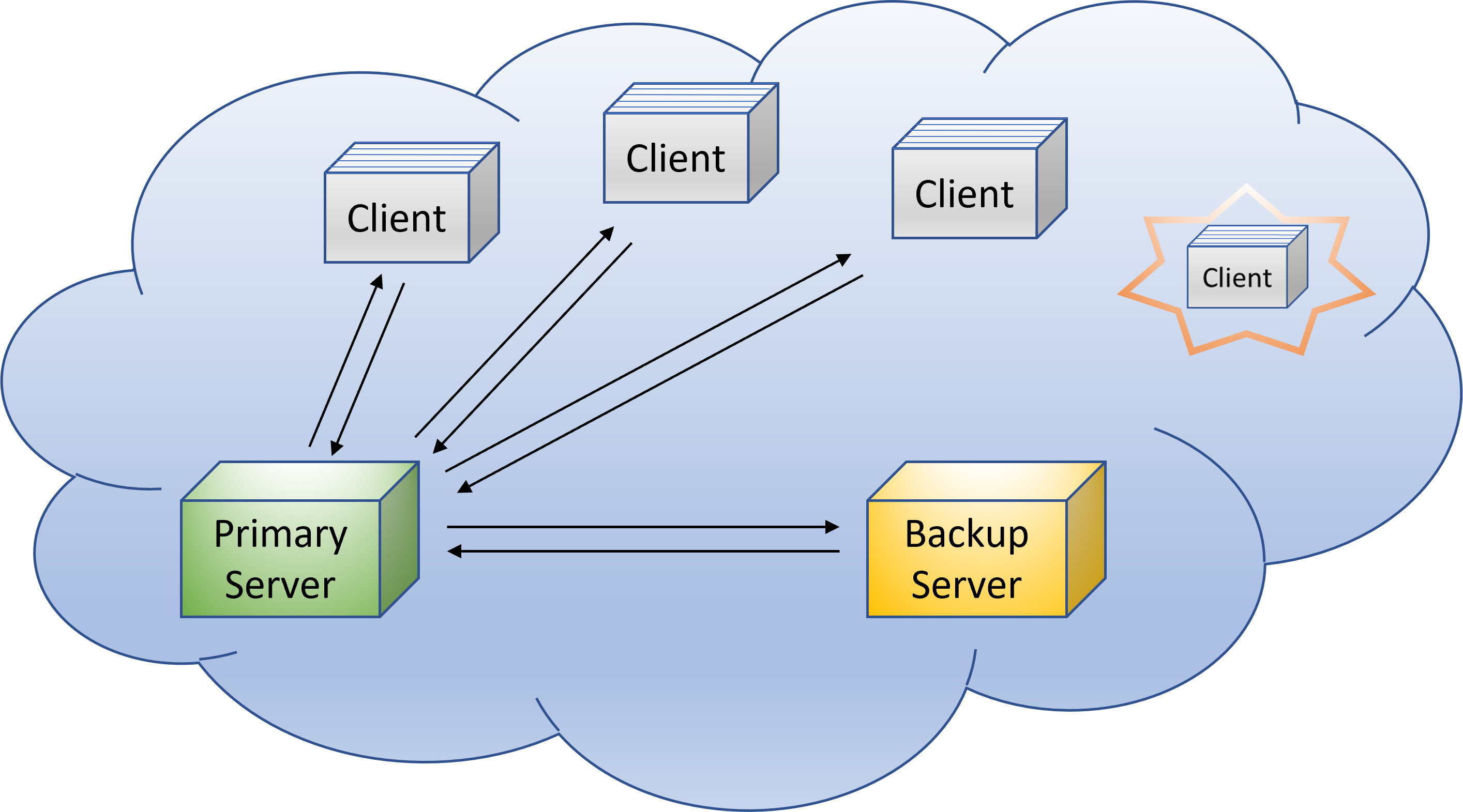}
    \caption{The overall architecture of ExpoCloud}
    \label{fig:arch}
\end{figure}

The overall architecture of ExpoCloud is shown in Figure~\ref{fig:arch}. It is a server-client architecture that uses a pull model to assign jobs to clients. Previous research \cite{Trellis} has shown suitability of such an architecture for distributed scientific computations. 

A distinguishing attribute of ExpoCloud is that it creates compute instances on-the-fly. Creating clients on-the-fly enables ExpoCloud to harness the great potential for large-scale concurrency provided by cloud-based compute engines. Creating replacement servers on-the-fly enables ExpoCloud to achieve effective fault tolerance.

Two features of the architecture are not shown in Figure~\ref{fig:arch}. First, there are two-way communication channels between the clients and the backup server. We detail the need for these channels in the section on fault tolerance below. Second, a client creates and manages {\em worker} processes. Each worker is responsible for executing a single task and communicating the results to the client.

The next section demonstrates how one can specify the example experiment described in the introduction. We will then show how the individual components of the architecture are implemented to provide time and budget efficiency as described in the introduction.

\subsection*{The example experiment}
To set up an experiment, one needs to write a short Python script that creates the primary server object, while providing it with the description of the tasks to be executed, the configuration of the compute engine and other optional arguments. 

We now show a possible script for exploring the parameter space when solving the agent assignment problem using B\&B. Here is the part of the script that constructs the list of tasks:

\small
\begin{lstlisting}[language=Python]
tasks = []
max_n_tasks = 50
n_instances_per_setting = 20

for options in [{Option.NO_CUTOFFS}, {}, {Option.HEURISTIC}]:
    for n_tasks in range(2, max_n_tasks + 1):
        for n_agents in range(n_tasks, 2 * n_tasks):
            instances = generate_instances(
                n_tasks, n_agents, 
                first_id = 0, 
                last_id = n_instances_per_setting - 1)
            for instance in instances:
                tasks.append(
                    Task(Algorithm(options, instance)))
\end{lstlisting}
\normalsize

The outer loop iterates over the three variants of the algorithm: the brute-force search, the classical B\&B search and the B\&B search with a heuristic. The following two nested loops iterate over the possible values of $n$ and $m$. For each parameter setting, a task is formed for each of the 20 generated problem instances. This task is added to the list {\tt tasks}.

The key component here is the {\tt Task} class, which the researcher needs to provide. For our experiment, this class may look as follows:

\small
\begin{lstlisting}[language=Python]
class Task(AbstractTask):
    def __init__(self, algorithm, timeout = 60):
        super(Task, self).__init__(algorithm, timeout)

    def parameter_titles(self):
        return self.instance.parameter_titles() + ("Options",)
        
    def parameters(self):
        return self.instance.parameters() + (set2str(self.options),)

    def hardness_parameters(self):
        def options2hardness(options):
            if Option.HEURISTIC in options: return 0
            if Option.NO_CUTOFFS in options: return 2
            return 1
                   
        return (
            options2hardness(self.options), 
            self.instance.n_tasks, 
            self.instance.n_agents)

    def result_titles(self):
        return self.algorithm.result_titles()

    def run(self):
        return self.algorithm.search()

    def group_parameter_titles(self):
        return filter_out(self.parameter_titles(), ('id',))
\end{lstlisting}
\normalsize

A brief description of each method follows:
\begin{itemize}
    \item {\tt parameter_titles} - returns the tuple of parameter names, which would appear as column titles in the formatted output. In the example implementation, these consist of the parameters of the problem instance, such as the number of agents and the number of tasks, appended by the parameters of the search algorithm being used.
    \item {\tt parameters} - returns the tuple of parameter values describing the task.
    \item {\tt hardness_parameters} - returns the subset of parameters used to determine the task's hardness. The default implementation in {\tt AbstractTask} says that task $T_1$ is as hard or harder than task $T_2$ if all the hardness parameters of the former are greater than or equal to the corresponding parameters of the latter. Note how the shown code converts the parameters of the search algorithm into a number, so as to adapt this default implementation. 
    
    Internally, the hardness of a task is stored as an instance of the {\tt Hardness} class defined inside {\tt AbstractTask}. The {\tt Task} class derives from {\tt AbstractTask} and may provide its own definition of {\tt Hardness}, thereby gaining full control over the way in which the hardnesses of two tasks are compared.
    \item {\tt result_titles} - returns the tuple of names of output quantities, such as the optimal time for executing the project and the time taken by the search algorithm. The actual tuple of output quantities is returned by the {\tt run} method described below.
    \item {\tt run} - executes the task by running the search algorithm to solve the problem instance. If the algorithm is implemented in Python, as in our example, the suitable method of the algorithm object is run. Otherwise, the algorithm can be run as a shell command.
    \item {\tt group_parameter_titles} - returns the tuple of parameter names that determine groups of tasks, as we now explain. Consider a state of the experiment, whereby results for three out of twenty problem instances for a particular setting of parameters have been computed. Suppose that a task timed out at this point, which disqualified the remaining sixteen tasks as being too hard. It stands to reason that the results for the three executed tasks should be discarded, since the average of the output quantities over only three tasks would have low statistical significance. On the other hand, had we obtained results for eighteen instances before a particularly hard task timed out, we may want to keep the results for this setting of parameters. 
    
    ExpoCloud makes the decision of whether to keep a parameter setting on a per-group basis. A group consists of all the tasks with the same values of the parameters returned by the {\tt group_parameter_titles} method.\footnote{This is somewhat similar to the idea of the {\tt GROUP BY} clause in SQL.} A group is kept when the number of solved tasks in the group is at least as large as the optional {\tt min_group_size} argument to the constructor of the server object. In the shown implementation, a group is defined by all the parameters besides the id of the problem instance within a particular setting of parameters.
    
    The default value of the {\tt min_group_size} argument is zero, which means that all the results are kept.
\end{itemize}

The next section of the script specifies the configuration for the compute engine and passes this configuration to the constructor of the engine object:

\small
\begin{lstlisting}[language=Python]
config = {
    'prefix': 'agent-assignment', 
    'project': 'bnb-agent-assignment',
    'zone': 'us-central1-a',
    'server_image': 'server-template',
    'client_image': 'client-template',
    'root_folder': '~/ExpoCloud',
    'project_folder': 'examples.agent_assignment'
}
engine = GCE(config)
\end{lstlisting}
\normalsize

The configuration is a dictionary with the following keys:

\begin{itemize}
    \item {\tt prefix} - the prefix used for the automatically generated names of compute instances. Several experiments with different prefixes may be conducted simultaneously.
    \item {\tt project} - the name identifying the project on the cloud platform.
    \item {\tt zone} - the zone to which the allocated compute instances will pertain. The current implementation of the GCE engine is limited to use a single zone. This limitation may be lifted in the future to enable an even larger scalability.
    \item {\tt server_image} and {\tt client_image} - the names of the machine images storing the configuration (such as the CPU family, the number of CPUs, the amount of RAM, etc) of all future server and client instances, respectively. An inexpensive configuration with one or two CPUs may be used for a server, while one may opt for 64 or more CPUs per instance for a client. ExpoCloud's clients make use of all the available CPUs automatically.
    \item {\tt root_folder} - the folder in which ExpoCloud resides on all the compute instances.
    \item {\tt project_folder} - the folder in which the user-provided scripts reside. The folder must be specified in the dotted format as shown in the listing.\footnote{This is the format in which the path must be specified when using the {\tt -m} command-line argument to {\tt python}.}  
\end{itemize}

In our case, the engine being used is the Google Compute Engine (GCE). Some dictionary keys for other engines may differ. For example, {\tt zone} is a GCE concept and a more suitable key name may be used in the extension class for another platform. %In the same vein, {\tt LocalEngine}, which is the engine class for running a local simulation of the experiment, is configured with a single value, namely the project folder.

Lastly, we construct the primary server object and call its {\tt run} method:
\small
\begin{lstlisting}[language=Python]
Server(tasks, engine).run()
\end{lstlisting}
\normalsize

% {\tt run_client} is a very short script, which imports the {\tt Task} class and runs the client object. For our experiment, it may look as follows:
% \small
% \begin{lstlisting}[language=Python]
% from src.client import Client
% from examples.agent_assignment.task import Task
% Client().run()
% \end{lstlisting}
% \normalsize

Once the experiment completes, the main ExpoCloud folder at the primary server will have an output folder containing a results file and a folder for each client instance. Such a client folder will contain files with the events sent by the client. ExpoCloud provides a script for convenient viewing of both the results and the client events related to the execution of the tasks.

ExpoCloud provides a local machine engine for running an experiment locally. The only change in the above script concerns the construction of the engine object:

\small
\begin{lstlisting}[language=Python]
engine=LocalEngine('examples.agent_assignment')
\end{lstlisting}
\normalsize

Once the experiment completes, the main ExpoCloud folder will have an output folder for each of the servers, as well as a file for {\tt stdout} and {\tt stderr} for each client. Running the experiment locally is useful both for small initial explorations. It also enables rapid development, since it makes it unnecessary to copy each updated version to the cloud and avoids the latencies associated with using the cloud.

\vspace{0.1in}
We now describe in detail how the primary server operates.

\subsection*{The primary server}
We first describe how the primary server stores the tasks, then outline the workings of the {\tt run} method at a high level, and lastly zoom in on the message-handling part of the primary server.

\subsubsection*{a. The tasks-related lists}
There are three tasks-related lists - the actual list of tasks and two auxiliary lists used for performance and fault tolerance. We describe them in turn.

The list of tasks, called {\tt tasks}, is sorted in the order of non-decreasing hardness of tasks. This order maximizes the number of tasks that are not attempted as a result of a previous task timing out. The original order of tasks is restored prior to the printing of results.

The list {\tt tasks_from_failed} consists of indices of the tasks that have been assigned to a client, but not completed due to a failure of the client instance. When a client requests tasks, the tasks in {\tt tasks_from_failed} are assigned first. The next task from {\tt tasks} is assigned only if {\tt tasks_from_failed} is empty.

Lastly, the list {\tt min_hard} consists of hardnesses of tasks that have timed out. Whenever the server is about to assign a task, it first checks whether the hardness of the task is equal or greater than any of the elements in {\tt min_hard}. {\tt min_hard} is kept small by only storing  the minimal elements. 

\subsubsection*{b. The {\tt run} method}
The primary server object's {\tt run} method executes an infinite loop. An iteration of this loop performs the following actions:
\begin{enumerate}
    \item Informs the backup server that the primary server is continuing to function properly. We refer to such a message as a {\em health update}.
    \item Handles handshake requests from newly started instances. The instance can be either a backup server or a client. We refer to the instance that has shaken hands with the primary server as an {\em active instance}. 
    
    In response to a handshake request, two-way communication channel with the instance is established.\footnote{Namely, the instance owns two queues registered with a {\tt SyncManager} object. The primary server creates the two corresponding queues at its end. {\tt SyncManager} is part of the {\tt multiprocessing} module of the Python standard library. It provides for low-latency communication, which makes the distributed approach effective even for fine-grained tasks.} In contrast to this channel, the queue for accepting handshakes is created by the primary server's constructor. When an instance is started, it gets the IP address of the primary server and the port number for handshake requests as command line arguments.
    
    In addition, if the instance is a client, a folder for storing the client events is created.
    
    \item Handles messages from clients. We outline the messages and how they are handled in the next section.
    \item Creates either the backup server or a new client instance. The creation of the backup server takes precedence. If the backup server is already running or the researcher opted to not use a backup server for the experiment, then a new client is created. Cloud compute engines do not let users to create instances in quick succession. Therefore, ExpoCloud uses exponentially increasing delays between attempts at creating cloud instances. 
    \item Terminates unhealthy instances. An active instance is unhealthy if it failed to send health updates to the server for the period of time specified by the {\tt HEALTH_UPDATE_LIMIT} constant. A non-active instance is unhealthy if it failed to shake hands with the primary server for the period of time specified by the {\tt INSTANCE_MAX_NON_ACTIVE_TIME} constant.
    \item Outputs the results once there are no tasks that have not been assigned to clients and all clients completed the tasks assigned to them.
\end{enumerate}

The servers do not stop once the results are output. Thus, the fault tolerance mechanisms continue to protect the results against a possible primary server instance failure.

\subsubsection*{c. The handling of messages}
The following is an outline of messages that may arrive to the primary server from the backup server and the client instances:
\begin{itemize}
    \item {\tt HEALTH_UPDATE} - the health update coming from either the backup server or a client. The primary server simply saves the timestamp of the last health update for each instance.
    \item {\tt REQUEST_TASKS} - the request for tasks by a client. The body of the message specifies the number of tasks requested. If there are remaining unassigned tasks, the {\tt GRANT_TASKS} message is sent in response. The body of this message contains the tasks assigned to the requesting client, including both the parameters and the full representation of the problem instances to be solved. If there are no unassigned tasks, the response is the {\tt NO_FURTHER_TASKS} message.
    \item {\tt RESULT} - the result of executing a task. The primary server stores the result with the task object.
    \item {\tt REPORT_HARD_TASK} - the report about a timed out task. The primary server updates the {\tt min_hard} list and sends the {\tt APPLY_DOMINO_EFFECT} message to all the clients, so they can terminate any task that is as hard or harder than the task just timed out.
    \item {\tt LOG} and {\tt EXCEPTION} - the report about an event related to executing a task or to an exception, respectively, sent by a client. The primary server stores the event in the fitting file corresponding to the client.
    \item {\tt BYE} - the client is done, which means that it had sent to the primary server the results for all the tasks assigned to it and had previously received the {\tt NO_FURTHER_TASKS} message. The primary server terminates the client instance, so the researcher will not incur any further charges. 
\end{itemize}

The primary server forwards a copy of each message from a client to the backup server. This keeps the backup server up-to-date and ready to take over should the primary server instance fail. This is further detailed in the section on fault tolerance below.

We will now describe the operation of the clients.

\subsection*{The clients}
We first describe the main loop of the client object's {\tt run} method, then detail how the workers are managed and lastly zoom in on the message-handling part of the client.

\subsubsection*{a. The main loop}
In contrast to the primary server, the client's main loop is not infinite -- it stops when there are no tasks assigned to the client, and no more tasks that can be assigned to it by the primary server (i.e. the {\tt NO_FURTHER_TASKS} message has been received).

Each iteration of the main loop performs the following actions:
\begin{enumerate}
    \item Sends the health update to the servers.
    \item Processes workers as detailed in the next section.
    \item Requests tasks from the primary server, subject to availability of idle CPUs and the {\tt NO_FURTHER_TASKS} message not having been received. Note that the tasks requested previously, but not yet granted are taken into account when determining how many idle CPUs there are. 
    \item Processes messages from the primary server as detailed in a separate section below. 
    \item If new tasks have been granted by the primary server, starts the worker processes to execute them.
\end{enumerate}

Foe each message sent to the primary server, the client sends a copy of the message to the backup server. The need for this is explained in the below section on fault tolerance. That section also details what the client does with the incoming messages from the backup server.

Once the main loop is exited, the client sends the {\tt BYE} message and completes.

\subsubsection*{b. The management of workers}
Each task is performed by a separate worker process. The client performs three action to manage the workers:
\begin{itemize}
    \item Processes messages arriving from the workers. A worker can send two messages - {\tt WORKER_STARTED} and {\tt WORKER_DONE}. In response to either message, the client sends the {\tt LOG} message with the corresponding body to the servers. The {\tt WORKER_DONE} message results in sending the {\tt RESULT} message as well.
    \item Takes accounting of the worker processes that are no longer alive (i.e. either done or terminated), so as to be able to assign the released CPUs to other tasks.
    \item Terminates processes whose tasks timed out. The client sends the {\tt REPORT_HARD_TASK} message to the servers for each timed out task.
\end{itemize}

\subsubsection*{c. The handling of messages from the primary server}
The following is an outline of messages that may arrive to the client from the primary server:
\begin{itemize}
    \item {\tt GRANT_TASKS} - one or more tasks have been assigned to the client. The task is added to {\tt tasks} list and a {\tt LOG} message is sent to the servers to record the event of the receipt.
    \item {\tt APPLY_DOMINO_EFFECT} - the hardness of a task that timed out is reported by the primary server. The client terminates all workers currently performing tasks of equal or greater hardness.
    \item {\tt NO_FURTHER_TASKS} - the primary server informs that there are no more tasks to be assigned. The client stores this information, so as to avoid requesting tasks and exit the main loop once all the worker processes are completed.
\end{itemize}

In addition to the above messages, there are the {\tt STOP}, {\tt RESUME} and {\tt SWAP_QUEUES} messages, used to achieve fault tolerance. These are detailed in the next section.

\subsection*{Fault tolerance}
One standard technique for achieving fault tolerance in a distributed system is by using redundancy \cite{Storm2012}. This is the approach we follow 
by employing a backup server that mirrors the primary one and substitutes for it in the case of a failure. A backup server is not used when the computation is performed using the local machine engine. As mentioned above, the researcher may choose to disable the use of the backup server. This may be desired for a short experiment.

We distinguish between three kinds of failure: client instance failure, backup server failure and primary server failure. Client failure does not require any special action besides registering the failure and re-assigning the tasks previously assigned to the failed client. The latter is achieved by maintaining the \mbox{{\tt tasks_from_failed}} list, as described above. In contrast, care needs to be taken to achieve correctness of recovery after a server failure. The following sections detail how this is achieved.

\subsubsection*{a. Creation of the backup server}\label{backup-creation}
The primary server creates the backup server in the same way as it creates clients. When a backup server instance does not yet exist, its creation takes precedence over the creation of a new client.

Note that the backup server maybe created either at the beginning of the computation or after a server failure.\footnote{We will see below how the case of primary server failure is reduced to the case of the backup server failure.} Therefore, we need to create the backup server under the assumption that the distributed computation is in progress.

To make sure that the newly created backup server is fully synchronized with the primary server, the primary server freezes its state prior to creating the backup server. First, it stops accepting handshake requests from new client instances. Second, it sends the {\tt STOP} message to the active clients, which causes them to refrain from  actions that may result in messages to the server. An exception is made for the health reports, which the clients continue to send.

Next, the primary server serializes its full state into a file in the output folder, creates a new instance on the compute engine, and copies the output folder to it. It then starts the backup server script on the new instance. This script unserializes the server object and runs its {\tt assume_backup_role} method. As the name suggests, this method converts the primary server object into a backup server one. First, it disconnects the server object from the clients' channels for communicating with the primary server and connects it to the channels for communicating with the backup server. Second, it creates a two-way channel for communicating with the primary server. Lastly, it shakes hands with the primary server, whereby two-way communication between them is established. Upon this handshake, the primary server resumes accepting handshake requests from new client instances and sends the {\tt RESUME} message to the clients, so they can resume normal operation. Finally, the backup server script starts the main event loop of the server object.

\subsubsection*{b. Primary and backup server coordination}
Whenever a new client shakes hands with the primary server, the latter sends the {\tt NEW_CLIENT} message to the backup server with the information about the client. In response to this message the backup server creates the client object and establishes communication with it. Similarly, whenever the primary server detects a client failure, it sends the {\tt CLIENT_TERMINATED} message to the backup server, which destroys the corresponding client object.

When a client sends a message to the primary server, it sends a copy of the message to the backup server. Thus, the backup server receives two copies of each message sent by the client. The copy received directly from the client is needed for the case when the primary server fails before forwarding the message to the backup server. The copy received from the primary server is needed to keep the two servers synchronized, as described below. 

The backup server takes actions based on the copy of the message received from the primary server. It simply pops the corresponding message received directly from the client off the queue. When the backup server registers the primary server failure, it will be ready to take over, with all the messages received directly from the clients after the last message forwarded by the primary server.

The backup server processes messages from clients in the same exact way as the primary server. It also sends messages to the clients that mirror the messages sent from the primary server. 

The mechanism of the backup server taking actions based on the copy of the message received from the primary server takes care of two possible causes of desynchronization. First, a client may fail after sending a message to the primary server, but before sending a copy to the backup server. Second, due to race conditions, it is possible for the two servers to handle messages from different clients, such as requests for tasks, in different order and end up in different states.

Similarly to how the backup server processes two copies of each message from a client, the clients processes two copies of each message from the servers - one received from the primary server and the other received from the backup server. A client performs actions only based on the messages received from the primary server and pops off the queue the corresponding messages received from the backup server. When the primary server fails, the remaining messages received from the backup server are treated as if they were from the primary server, as detailed in the next section.

\subsubsection*{c. Handling server failure}
In response to the backup server failure, the primary server simply creates the new backup server as outlined in the last section.

In the case of the primary server failure, the backup server changes its own role to being the primary server. It then proceeds to create a temporary connection to each client's inbound queue for communication with the primary server and sends a {\tt SWAP} message. In response to this message, the client swaps the queues for communicating with the primary server with those for communicating with the backup server. After this, the client is ready to proceed normally. 

Thus, the case of the primary server failure is now reduced to the case of the backup server failure discussed above.

%%The above relatively simple approach effectively handles the most common failures. Several edge cases that need to be specially handled are the subject of the next section.
%%\subsubsection{Special cases}

One special case is when the primary server fails after creating a client instance, but before the new client shakes hands and the backup server is updated. In this case, there is a dangling client instance incurring charges for the researcher. To avoid this, as part of the backup server assuming the role of the primary server, it requests from the engine the list of all compute instances and deletes all client instances that are not represented by an existing client object.

%\section*{Results}
%% Results should be clear and concise.

%\section*{Discussion}
%% This should explore the significance of the results of the work, not repeat them. A combined Results and Discussion section is often appropriate. Avoid extensive citations and discussion of published literature.

%\section*{Conclusions}
%% The main conclusions of the study may be presented in a short Conclusions section, which may stand alone or form a subsection of a Discussion or Results and Discussion section.

\section*{Discussion and conclusions}
We have presented the ExpoCloud framework for distributed computing using cloud compute engines. Unlike the existing tools geared towards business workloads \cite{burns2016borg}, ExpoCloud is specifically designed to make it easy to execute large parameter-space explorations. It addresses the four main concerns of the researcher: ease of defining the workload, harnessing as much compute power as can be used to speed up the experiment, eliminating computations that do not or are unlikely to complete in a reasonable amount of time, and avoiding unnecessary charges. Combined with mechanisms for fault tolerance, these properties make ExpoCloud a fitting tool for many research projects requiring large-scale parameter-space explorations. Future work may consider executing workloads with task dependencies, integrating ExpoCloud with existing tools, and addressing security concerns. 

\section*{Acknowledgements}
Access to the Google Compute Engine was provided through the Israel Data Science Initiative.

\bibliography{main}

\end{document}